\documentclass{ws-jin}
\usepackage{url,wsnatbib}
\begin{document}

\markboth{V. Salari et.al}{On the Classical Vibrational Coherence of Carbonyl Groups ...}

\catchline{}{}{}{}{}

\title{On the Classical Vibrational Coherence of Carbonyl Groups in the Selectivity Filter Backbone of the KcsA Ion Channel}

\author{V. Salari}

\address{Department of Physics, Isfahan University of Technology\\
Isfahan 84156-83111, Iran\\Foundations of Physics Group, School of Physics, Institute for Research in Fundamental Sciences (IPM), Tehran 19395-5531, Iran\\
\email{vahidsalari@cc.iut.ac.ir}
}

\author{M. Sajadi}

\address{Department of Physics, University of Shahrekord\\
Shahrekord, 88186/3414, Iran
}

\author{H.Bassereh}

\address{Department of Physics, Isfahan University of Technology\\
Isfahan 84156-83111, Iran\\
\email{}
}

\author{V. Rezania}

\address{Department of Physical Sciences, Grant MacEwan University\\
Edmonton T5J 4S2, Canada\\
\email{}
}

\author{M. Alaei}

\address{Department of Physics, Isfahan University of Technology\\
Isfahan 84156-83111, Iran\\
\email{}
}

\author{J. A. Tuszynski}

\address{Department of Physics and Experimental Oncology, University of Alberta\\
Alberta T6G 2J1, Canada\\
\email{jackt@ualberta.ca}
}

\maketitle

\begin{history}
\received{Day Month Year}
\revised{Day Month Year}
\end{history}

\begin{abstract}
It has been suggested that quantum coherence in the selectivity filter of ion channel may play a key role in fast conduction and selectivity of ions. However, it has not been clearly elucidated yet why classical coherence is not sufficient for this purpose. In this paper, we investigate the classical vibrational coherence between carbonyl groups oscillations in the selectivity filter of KcsA ion channels based on the data obtained from molecular dynamics simulations. Our results show that classical coherence plays no effective role in fast ionic conduction. 
\end{abstract}


\section{Introduction}

Since the discovery of potassium $K^+$ channels, a surge of interest among investigators has been aroused to understand and explain their fascinating molecular mechanisms of action.  These protein complexes assemble from several proteins creating circular pores through the membrane and operate in a fast and precise manner. The selectivity filter is believed to be responsible for the selection and fast conduction of particular ions across the membrane of an excitable cell. Other (generally larger) parts of the molecule such as the pore-domain gate control the access of ions to the channel protein. The KcsA channel, a potassium channel from bacterium Streptomyces lividans conducts $K^+$ ions with a high efflux rate ( $10^7$ - $10^8$ ions per second \citep{MacKinnon}) while selecting $K^+$ over $Na^+$ with a remarkable rate of $10^4$ to 1\citep{Doyle}. This, of course, should be particularly appreciated by noting that the difference in atomic radii of $K^+$ and $Na^+$ is about 0.38 nm.

Following the determination of an atomic resolution structure of the bacterial KcsA channel by MacKinnon and co-authors \citep{MacKinnon}, it became increasingly apparent that classical models such as the 'snug-fit' model \citep{Noskov}, might not be adequate enough to explain such a delicate selectivity process. Furthermore, the question was raised whether the selectivity filter can also adopt non-conducting states, acting as a 'filter-gate' and how these states could become synchronized with the mechanical opening and closing of the pore-domain gate facilitating conduction. In particular, it was found that 'selectivity' can be attributed to the specific Coulomb-coordination geometry between the ion and the surrounding dipolar carbonyl ligands lining the so-called P-loop domain of the filter region \citep{Bostick,Berneche}.




\subsection{The Physics inside the Selectivity Filter}
The 3.4 nanometer long KcsAchannel is comprised of a 1.2 nanometer long selectivity filter that is composed of four P-loop strands and whose structure is similar to that of the alpha-helix. The difference is due to the fact that they do not have hydrogen bonds which connect conformons to each other in an alpha-helix structure. Each P-loop is composed of five amino acids: [T(Theronine, Thr75), V(Valine, Val76), G(Glycine, Gly77), Y(Tyrosine, Tyr78), G(Glycine, Gly79)] linked by peptide units (H-N-C=O) in which N-C=O is an amide group and C=O is a carbonyl group. Carbonyls are responsible fortrapping and displacement of the ions in the filter.
\begin{figure}
\includegraphics[height=5cm,width=12cm]{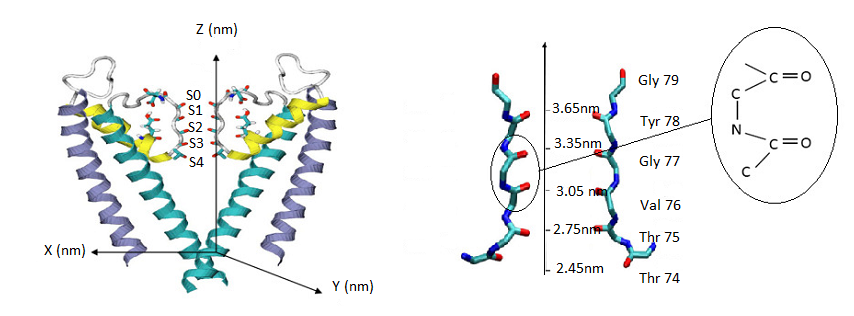} 
     \caption{(Left side) A representation of KcsA ion channel indicating the trapping sites of ions, which are labeled as S0, S1, S2, S3 and S4.  Center) Two KcsA-P-loop strands in the selectivity filter, composed of the sequences of TVGYG amino acids [T(Theronine, Thr75), V(Valine, Val76), G(Glycine, Gly77), Y(Tyrosine, Tyr78), G(Glycine, Gly79)] linked by peptide units H-N-C=O. Right side) Representation of carbonyl (C=O) groups in the selectivity filter. The whole peptide units are not shown here.}
\label{fig1}
\end{figure}

The movement of ions through a selectivity filter in the ion channel pore is affected by important physical processes like ion-ion, ion-water, ion-backbone, and ion-electric-field interactions. 
The most important reason for the movement of biomolecules is due to the action of intermolecular forces. The main forces are the Coulomb interactions between dipoles and ions. For example, the dipole moment of a water molecule is $P_W=6.13\times10^{ - 30} C.m$, so the potential energy between two water molecules is 1.25eV, the potential energy involving a water molecule and a potassium ion is about 2.4eV and involving two potassium ions is 9.6eV \citep{Duane}. In general, a comparison between the strengths of the various potential energy types reveals the following relationship: $E_{ion-ion}\succ E_{ion-dipole}\succ E_{dipole-dipole}$. The ion-ion potential energy is proportional to $1/r$, ion-dipole to $1/r^2$ , and dipole-dipole to $1/r^3$. 
We note that this could be in fact due to the interactions between ions and carbonyl groups in a selectivity filter, which mediate this Coulomb interaction. Each ion is bound to 8 carbonyl groups in a selectivity filter. Carbonyl groups possess electrical dipoles with dipole moments $P_{c = o}  = 7.2 \times 10^{ - 30} C.m$    \citep{Chung}.   As a result, the orientation of carbonyl groups would change relative to the movements of ions.  MD simulations have shown that the C=O bond in a carbonyl group may oscillate radially and angularly with typical amplitudes of $a_n^{c = o,r}  = 0.005A^ \circ $
 and $a_n^{c = o,\theta }  = 0.01A^ \circ $ , respectively \citep{Bucher}. If the dipole moment of carbonyl is not oriented toward the ion, it experiences a torque which forces it to return to a stable equilibrium. A carbonyl group generally orients itself to have its more electronegative oxygen atom near the carbon atom.  However, when an ion gets very close to a carbonyl group, it forces the C=O to align its dipole with the electric field. Moreover, the dipole-dipole interactions between adjacent carbonyls cause a longitudinal oscillation through the strand. The amplitude of longitudinal oscillation varies in the range $a_n^{Si-Sj }  = 3-10\times 10^{ - 3}A^ \circ $  \citep{Bucher}. The resulting Coulomb interaction with the innermost filter-ion (corresponding to a field strength of up to $ 9\times 10^{ 9} V.m^{-1}$ ) destabilizes the carbonyl binding pockets within the filter configuration, eliminates the âlocalityâ of energy profiles and finally leads to the ejection of the outermost ion from the channel \citep{Berneche}.

\subsection{Quantum Coherence in the Selectivity filter backbone}
The structural X-ray crystallography and MD simulations of $K^+$ ion-channels revealed a picosecond transition time that raised the possibility of the occurrence of quantum coherence in these systems as demonstrated in other biological processes such as energy transfer in alpha-helix structures \citep{Asadian}. Plenio and his colleagues have been attempting to determine whether there exists such quantum effects in ion channels or not \citep{Vaziri, Plenio, Tsomokos, Caruso}. Recently, Vaziri et al. \citep{Vaziri} and Ganim et al. \citep{Ganim} proposed the presence of quantum coherence by arguing that the backbone structure of the selectivity filter is not rigid as expected in classical models and studied vibrational excitations in  $K^+$  ion-channels. They discussed a possible emergence of resonances at the picoseconds (ps) scale in t he backbone amide groups that can play a role in mediating ion-conduction and ion-selectivity in the selectivity filter. Summhammer et.al \citep{Summhammer} also investigated the interaction of a single potassium ion within the surrounding carbonyl dipoles by analyzing solutions of the Schroedinger equation for the bacterial KcsA ion-channel.  They showed that alkali ions can become highly delocalized in the filter region at sufficiently high temperatures. They claimed that a quantum mechanical calculation is needed to explain a fundamental biological property such as ion-selectivity in trans-membrane ion-channels. 
It has been recently shown that excitation energy transfer along a P-loop strand cannot be fully classical but it should be a quantum-classical hybrid to make the excitation energy transfer more efficient and faster \citep{Bassereh}. If quantum effects do play a critical role in filter-ion coordination, it is feasible that these delicate interactions could leave their quantum traces in the overall conformation and the molecular gating state of the entire protein
 \citep{Moradi, Summhammer, Salari}. Nevertheless, so far it has not been conclusively established why classical coherence is not as important as quantum coherence in ion channels. Here, we investigate this in detail and discuss whether a classical coherence can play a role to help for fast conduction of ions in the selectivity filter of ion channel. Here, we would like to investigate the effect of classical vibrational coherence on ion transport in the selectivity filter backbone.

\section{Method}
Our molecular dynamics (MD) simulations are based on a model of the KcsA channel (Protein Data Bank, 1K4C.pdb), embedded in a palmitoyloleoyl phosphatidylcholine (POPC) lipid bilayer. The system was built from a cubic box of a 7.8 nm side that comprises KcsA (four subunits of 97 amino acids, 5292 atoms), water molecules (TIP3P model, 42296 atoms), 3K and 2K in the pore and 12 CL in the bulk (the entire system is electrically neutral). The AMBER 03 force field parameters \citep{Cornell} and GROMACS 4.5.3 software \citep{Hess} was employed to perform two simulations: once for the time step of 1 ps during 10 ns and once for the time step of 1 fs during 10 ps. The protein was equilibrated in (N, V, T) then (N, P, T) ensembles. The temperature was kept at 300 K by Nose-Hoover coupling algorithm and the pressure was kept at 1 bar by Parrinello-Rahman coupling algorithm. The system is oriented along the z-axis. A cutoff was used for long-range interactions, namely: 0.12 nm forthe van der Waals interaction and 0.14 nm for electrostatic interactions. Using  the Particle-Mesh Ewald  (PME)  method,  the  electrostatic  interactions  are  calculated.We have used the following abbreviations for amino acid identification: GLY1=GLY79, TYR=TYR78, GLY2=GLY77, VAL=VAL76, THR2=THR75, THR1=THR75 (i.e. hydroxyl). Each carbonyl group is a C=O compound in which the vibrations of C and O atoms are investigated separately.

\section{MD Simulation of Amino Acid Vibrations}
\begin{figure}
\includegraphics[height=8cm,width=14cm]{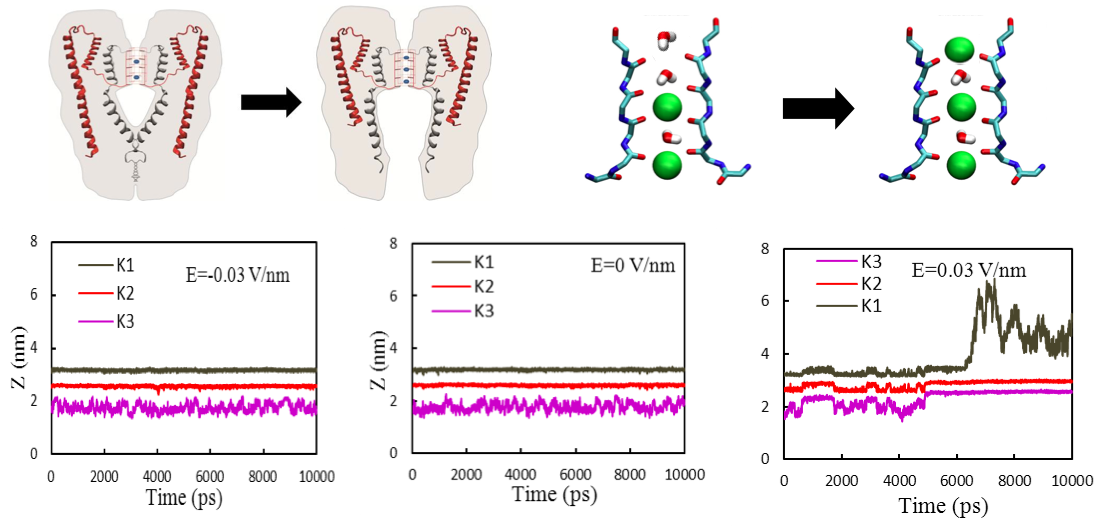} \caption{K1, K2 and K3 are three potassium ions in a KcsA ion channel where K1 is located in the site $S_0$, K2 located in $S_2$ and K3 in $S_3$. When the membrane potential is positive and high enough (i.e. an electric field equal to 0.03 V/nm), the ion K1 jumps to the outside of the SF and the flow of ions is started. For similar magnitude but negative electric potential, ions are trapped at specific locations and thus, there is no flow which indicates the unidirectional nature of the KcsA channel for positive electric potential. Trapping of ions is demonstrated for zero potential, too.}
\label{fig2}
\end{figure}

In order to demonstrate the dynamics of KcsA ion channels we performed a simulation. At first, we intend to investigate the unidirectional nature of the channel in regards to ion flow.  A fairly strong constant electric field ({$\sim$ .03 V/nm}) is applied once in upward and once in downward direction. Fig~\ref{fig2} shows how the third ion (i.e. K1 in the site $S_0$) jumps outside of the filter (after about 7 nano seconds) when the potential difference of the membrane is positive(i.e., V=100 mV, equal to the electric field E=0.03 V/nm). The other two ions, K2 and K3, are located in sites S2 and S4 respectively. Since the channel's performance and also the flow of ions are dependent on the natural potential of cell's membrane, imposing different fields from the natural field of the two sides of the membrane changes this performance. Imposing different constant fields, we perceived that the flow of Potassium ions from the ion channel is only performed in its natural stable potential and changing this imposed field causes stop in the movements of potassium ion.\\

 Here, we have analyzed the vibrations of carbonyl groups (C=O), or more precisely, the vibrations of oxygen (O) and carbon (C) atoms, for a negative (i.e., -70 mV) and positive (i.e., +30 mV) membrane electric potential difference. 
Our attempt is intended to address the question whether a motional coherence of carbonyl vibrations helps in fast conduction of ions or not. We consider the two states of the channel: open and closed. We have clearly a current for an open channel and thus we have to expect more coherence between carbonyl vibrations to help in the fast flow of ions. Conversely, we should expect less coherence for a closed channel in which the ions are almost trapped and exhibit no significant motion. We have considered a -70 mV state for the closed channel and a +30 mV state for the open channel. Two ions are located in S2 and S4 for the closed state in our simulations and three ions are located in S0, S2 and S4 for the open state (as the same as the configuration in Fig~\ref{fig2}).

\subsection{Open channel}
\begin{figure*}
\includegraphics[height=5cm,width=6.5cm]{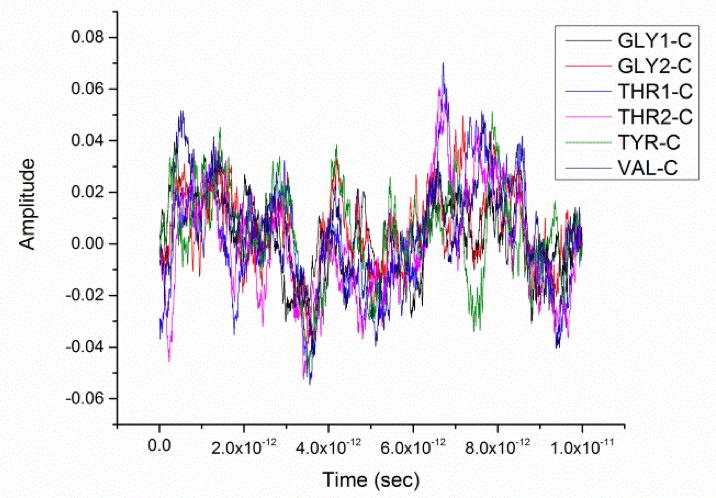}
\includegraphics[height=5cm,width=6.5cm]{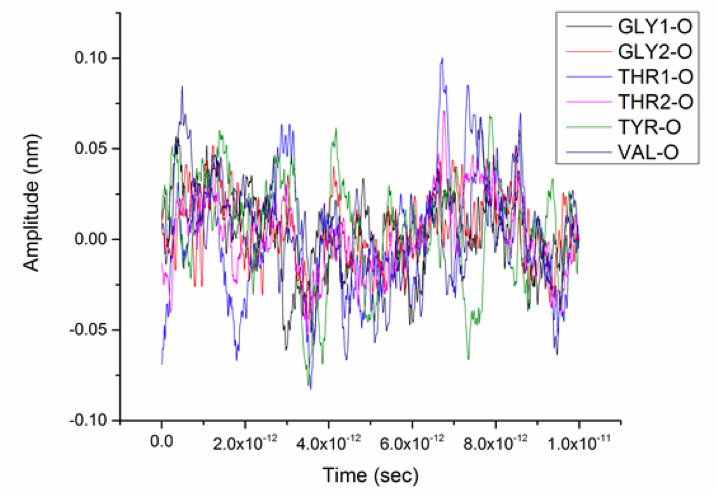}
 \caption{The oscillations of C atoms of the carbonyl groups in a 10 ps time interval with a 1 fs time step for the open state of the ion channel when there are three ions in the selectivity filter. Right) The oscillations of O atoms of the carbonyl groups in a 10 ps time interval with a 1 fs time step for the open state of the ion channel when there are three ions in the selectivity filter. It is seen that the carbon atoms vibrate with amplitudes maximally to 0.06 nm but the amplitudes of oxygen atoms vary up to 0.1 nm}
\label{fig3}
\end{figure*}

Based on our MD simulation, we have shown separately the fluctuations of C and O atoms of carbonyls as a function of time(see Fig ~\ref{fig3}). The amplitudes are calculated by decreasing the average of the data from the original data obtained by MD to avoid DC offset in the diagrams. On the other hand, the vibration amplitude of each atom is $r-r_0$ where refers to the original data and $r_0$ is the average of all these obtained data points. Fig~\ref{fig3} demonstrates that the carbon atoms vibrate with amplitudes maximally to 0.06 nm but the amplitudes of oxygen atoms vary up to 0.1 nm.\\

\begin{figure*}
\includegraphics[height=5cm,width=6.5cm]{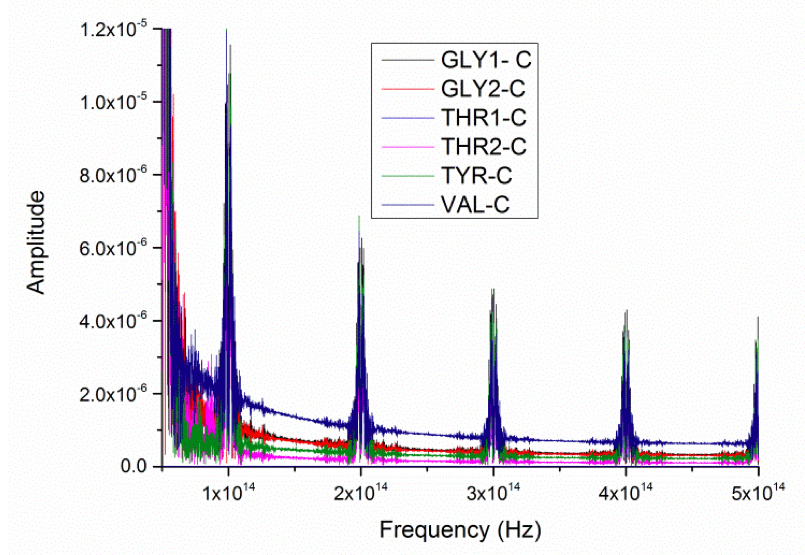} 
\includegraphics[height=5cm,width=6.5cm]{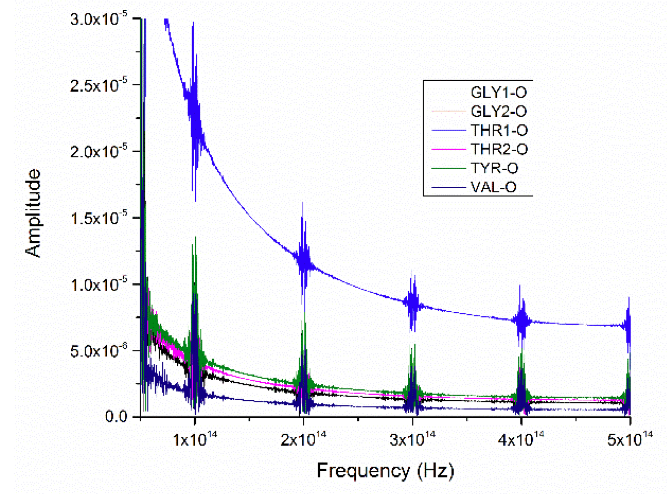}
\caption{Vibrational frequencies of C and O atoms of the carbonyl groups are obtained for an open state in +30 mV based on the FFT method. It is seen that there are some harmonies between the vibrations of C and O atoms. }
\label{fig4}
\end{figure*}

We obtain the frequencies of vibrations for C and O atoms via fast Fourier transform (FFT) method. The frequencies are plotted in Fig~\ref{fig4} in the range 50-500 GHz using the Origin software.\\ 

\subsection{Closed channel}

\begin{figure*}
\includegraphics[height=5cm,width=6.5cm]{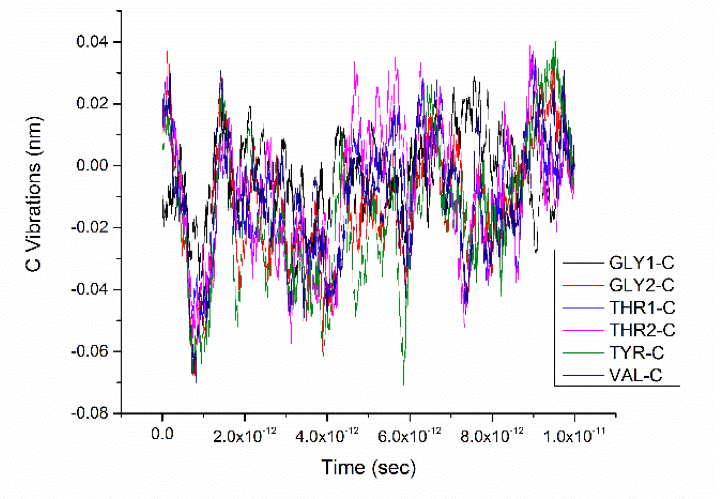}
\includegraphics[height=5cm,width=6.5cm]{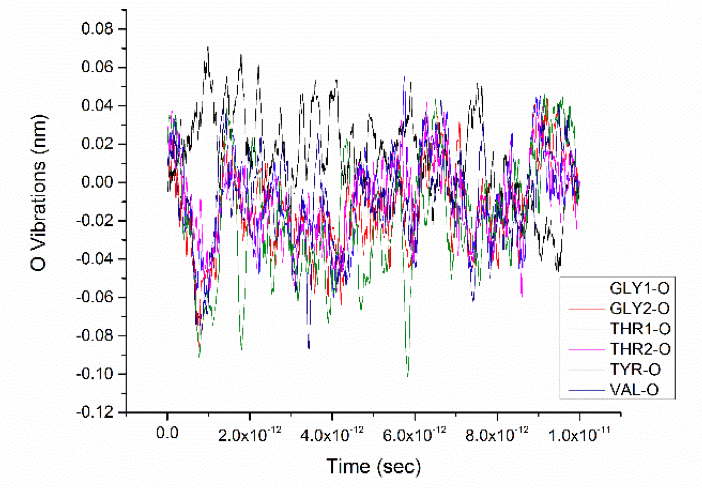}
 \caption{The oscillations of C  and O atoms of the carbonyl groups in a 10 ps time interval with a 1 fs time step for the closed state of the ion channel when there are two ions in the selectivity filter. It is seen that the carbon atoms vibrate with amplitudes maximally to 0.06 nm but the amplitudes of oxygen atoms vary up to 0.1 nm}
\label{fig5}
\end{figure*}

\begin{figure*}
\includegraphics[height=5cm,width=6.5cm]{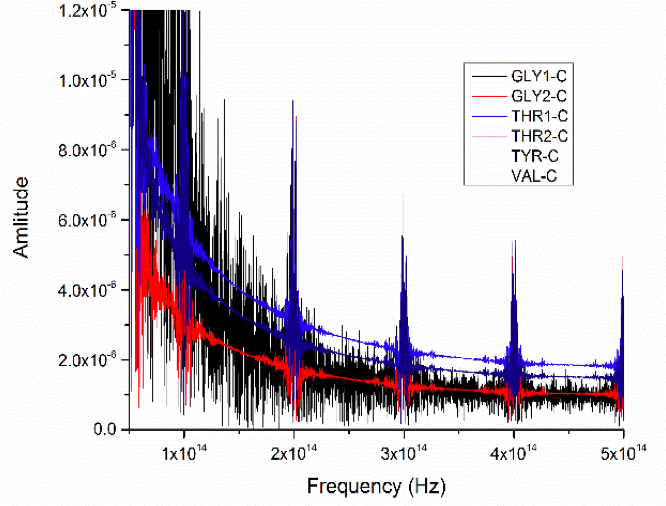} 
\includegraphics[height=5cm,width=6.5cm]{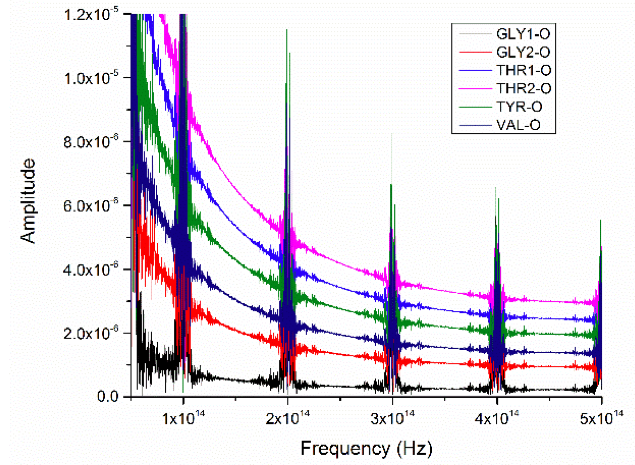} 
\caption{Vibrational frequencies of C and O atoms of the carbonyl groups are obtained for a closed state in -70 mV based on the FFT method. It is seen that there are some harmonies between the vibrations of C atoms and O atoms, respectively.}
\label{fig6}
\end{figure*}

Now, we repeat our previous analysis, this time for the closed state of the channel (i.e., V= -70 mV) in which there are two ions in the selectivity filter. The fluctuations of C and O atoms are plotted in the Fig~\ref{fig5}, in which it is seen again that the carbon atoms vibrate with amplitudes maximally to 0.06 nm but the amplitudes of oxygen atoms vary up to 0.1 nm.\\
Now, we obtain frequencies via the FFT method again. The results for C  and O atoms are plotted in Fig~\ref{fig6}. \\

It is seen in figures ~\ref{fig4} and ~\ref{fig6} that there are some harmonies (or coherence) between the vibrations of C and O atoms. Now, we would like to compare the degrees of coherence for open and closed states to see whether the classical coherence plays any role in ion transport in the filter or not. 

\section{Correlation and Degree of Coherence}
Based on the Wiener-Khinchine theorem \citep{Wiener, Nahin} the power spectrum  $P(\omega)$ of a signal $S(t)$ is the Fourier transform of its autocorrelation function $A(\tau)$  for nonzero $\tau$. If $A(\tau)=0$, then the signal is said to be "completely uncorrelated", then $A(\tau)$, is a Dirac delta function and its corresponding power spectrum is flat, or "white", which is completely uncorrelated, while a signal with a "colored" spectrum has some amount of correlation in it. The cross-correlation function $Cr_{xy}$ is a similar function to autocorrelation function but for correlation between two signals, $x$ and $y$. The cross-correlation function is a quantitative operation in the time domain to describe the relationship between data measured at a point and data obtained at another observation point \citep{Broch}. The cross-correlation function is

\begin{equation}
Cr_{xy} (\tau ) =  < {S_x (t).S_y (t + \tau )>} 
\end{equation}
Where  $S_x(t) $ is the magnitude of the signal $x$ at time $t$, and $S_y (t + \tau )$ is the magnitude of the signal y at a time $\tau$ later. The cross-spectral density function,  $P_{xy} (\omega )$, can be obtained by applying the Fourier transform to the cross-correlation function.

\subsection{Pearson Correlation}
Correlation between sets of data is a measure of how well they are related. Correlation coefficient formulas are used to find how strong a relationship is between data. The correlation coefficient ranges from -1 to 1. A value of 1 implies a strong positive relationship between X and Y. A value of -1 implies a strong negative relationship where the all data points lie on a line for which Y decreases as X increases. A value of 0 implies indicates no relationship at all and there is no linear correlation between the variables. The most common measure of correlation is the Pearson Correlation\citep{Aldrich}. The correlation in the same direction is called positive correlation. For example, if one variable increases the other also increases, and if one variable decreases the other also decreases. On the other hand, the correlation in the opposite direction is called negative correlation, in which if one variable increases the other decreases and vice versa. There is no relationship between the two variables if the value of one variable changes and the other variable remains constant, thus this is called no (or zero) correlation.

\begin{figure*}
\includegraphics[height=5cm,width=6.5cm]{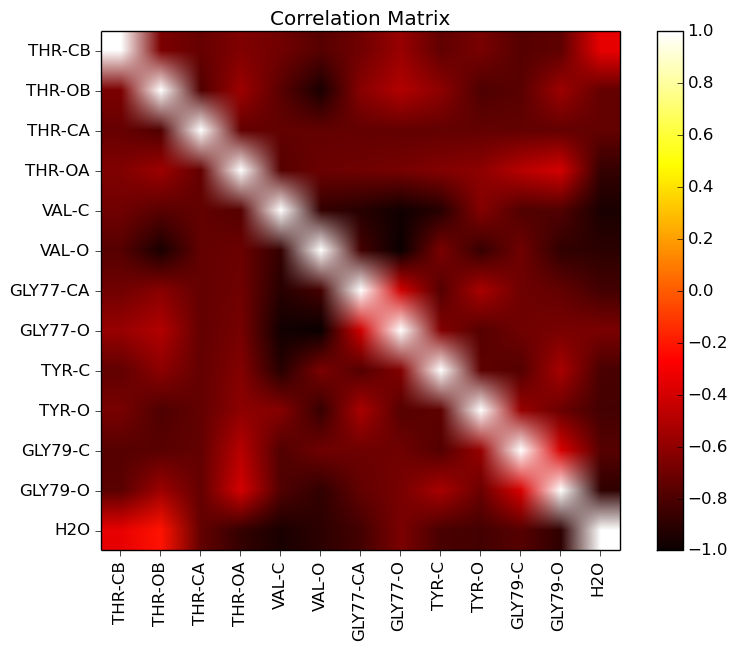}
\includegraphics[height=5cm,width=6.5cm]{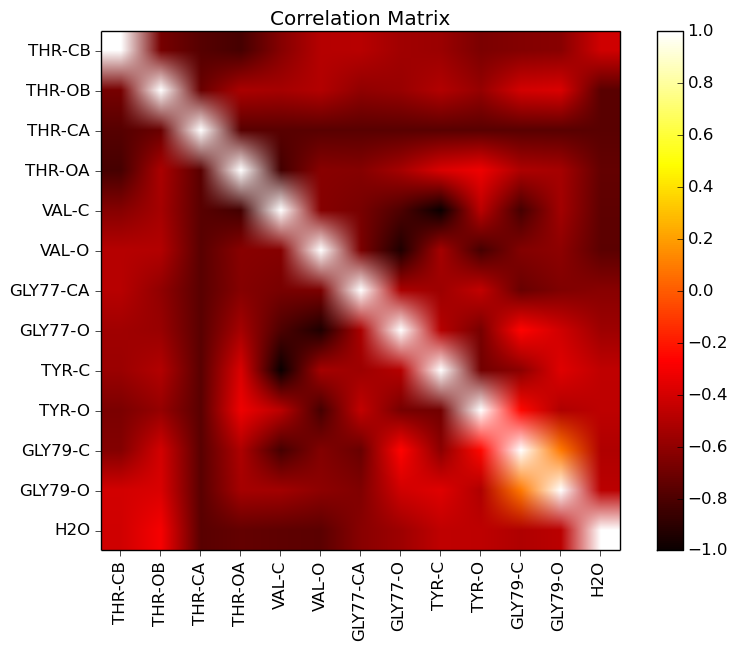}\\
\includegraphics[height=5cm,width=6.5cm]{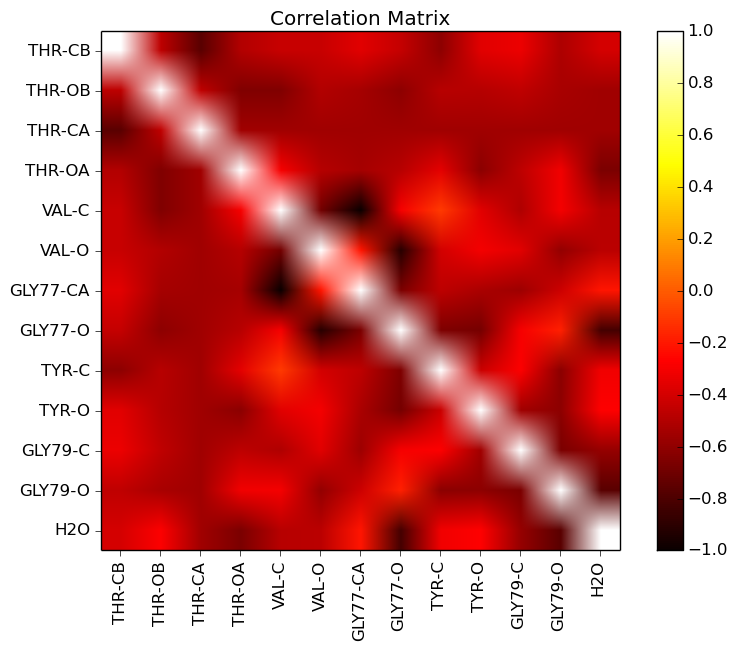}
\includegraphics[height=5cm,width=6.5cm]{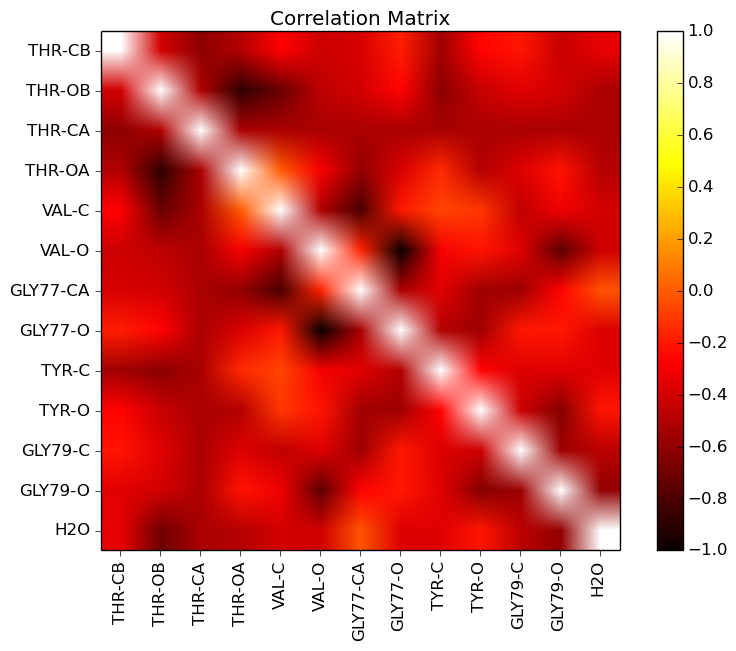}
 \caption{Pearson correlation plots of amino acids for different voltages, V=-100 mV (top, left), -70 mV(top, right), 30 mV(bottom, left), and 100 mV (bottom, right). The vibrational correlations for carbonyl oscillations in the closed states (-100 mV and -70 mV) are higher than the open states (+30 mV and +100 mV), since the black points, which are indicating higher correlation (while negative), have occupied more area of the plots in the closed states. The bright colours have lower intensities in the all figures.}
\label{fig7}
\end{figure*}

By using the Pearson correlation for the data in figures ~\ref{fig3} and ~\ref{fig5} as well as the data for -100 mV and +100 mV (data are not shown) we have plotted the Pearson correlation via Python software for each voltage in Fig ~\ref{fig7}, which shows that the correlation decreases by increasing voltage. In other hand, the vibrational correlations for carbonyl oscillations in the closed states are higher than the open states. In fact, the higher negative correlations are seen obviously in the closed states because the black points, which are indicating higher correlation (while negative), have occupied more area of the plots in the closed states.

\subsection{Degree of Coherence}
\begin{figure}
\includegraphics[height=5cm,width=6.5cm]{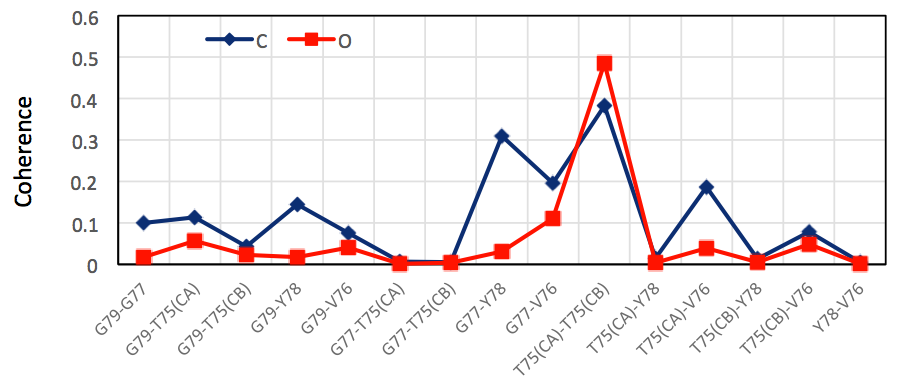}
\includegraphics[height=5cm,width=6.5cm]{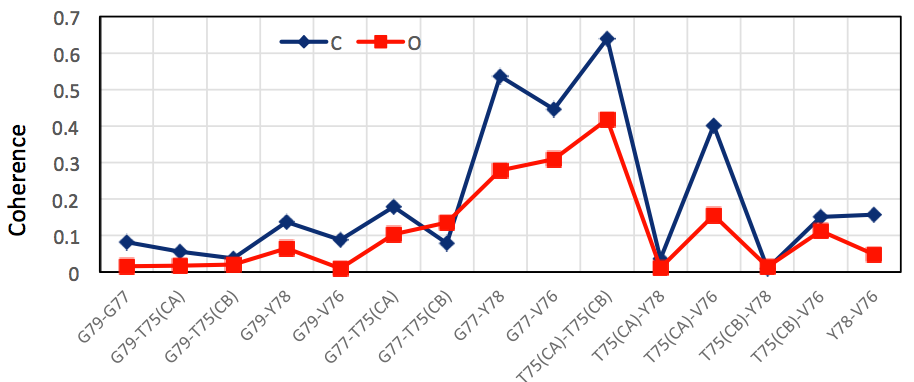}
 \caption{The degree of coherence between the vibrations of C and O atoms of the carbonyl groups in the presence of three K ions in the filter for the potential +30 mV. Right) The degree of coherence of C and O vibrations for the closed state of the channel in -70 mV. It is seen for the open state (i.e. +30 mV) that the highest degree of coherence for C atoms is 0.4 and the major coherence varies in the range 0-0.2 which is a weak partial coherence. The maximum degree of coherence for O atoms is 0.5 which only happens between two O atoms, while for the rest of O atoms the major degree of coherence is in the range 0-0.05 which signifies a very weak partial coherence (or almost no coherence) between O atoms in the open state of the channel. On the other side, it is clearly seen for the closed state (i.e. -70 mV) that there are degrees of coherence for C atoms vibrations even up to 0.7 and four mutual degrees of coherences between 0.4 and 0.7, which indicates that there is a rather strong partial coherence between the C vibrations.}
\label{fig8}
\end{figure}

Now, we should establish the amount of harmony between the vibrations numerically by using the degree of coherence. In order to do this,  we obtain and plot the degree of coherence between carbons (or oxygens) of amino acids two by two based on a standard coherence procedure in signal processing, i.e.$C_{xy}  = \frac{{\left| {P_{xy} } \right|^2 }}{{P_{xx} P_{yy} }} $,  in which $C_{xy}$is the degree of coherence between two signals $S_x(t)$ and $S_y(t)$, $P_{xy}$ is the cross spectral density between the signals, and $P_{xx}$and $P_{yy}$ are the autospectral density of $S_x(t)$ and $S_y(t)$ signals, respectively \citep{Bendat, Stoica}. If$ C_{xy}=1$, this corresponds to complete coherence, while if $0 \prec C_{xy}  \prec 1$   it is a partial coherence and if $C_{xy}=0$ there is no coherence. \\
The average degree of coherence between C and O atoms for both open and closed states of the channel are plotted in Fig~\ref{fig8}. It is seen for the open state (i.e. +30 mV) that the highest degree of coherence for C atoms is 0.4 and the major coherence varies in the range 0-0.2 which is a weak partial coherence. The maximum degree of coherence for O atoms is 0.5 which only happens between two O atoms, while for the rest of O atoms the major degree of coherence is in the range 0-0.05 which signifies a very weak partial coherence (or almost no coherence) between O atoms in the open state of the channel. On the other side, it is clearly seen for the closed state (i.e. -70 mV) that there are degrees of coherence for C atoms vibrations even up to 0.7 and four mutual degrees of coherences between 0.4 and 0.7, which indicates that there is a rather strong partial coherence between the C vibrations. On the other hand, the degrees of coherence for O atoms are lower than C atoms for the closed state, but much stronger than O atoms for the open state.

\section{conclusion}
In this paper, we have investigated the classical motional coherence and correlation between carbonyl groups in the selectivity filter of KcsA ion channels via molecular dynamics to determine whether the classical coherence plays any significant role in fast conduction of ions through the filter. We have performed our simulations for two states of an ion channel. The first state is an open state in which there are three ions in the filter and the electric potential of membrane is +30 mV and thus there is current flowing through the channel. The other state is the closed state of the channel in which there are two ions in the channel and the electric potential of membrane is -70 mV and there is no current. We have observed that the degrees of coherence between the vibrations of the selectivity filter backbone, which experiences the ions for the closed state of the channel, are much higher than the degrees of coherence of vibrations in the open state of the channel. We also have tried Pearson correlation for other potential of membranes, -70 mV and -100 mV for closed state and +30 mV and +100 mV for open state, and again we observed that correlation between the vibrations for closed state is higher than the open state. This observation confirms that the classical coherence vibrations of carbonyl groups play no major role in the fast conduction of ions. The most important role is due to the positive electric field produced by changing the concentration of ions at both sides of the membrane. Yet, quantum excitation energy transfer or quantum coherence may be still affective for such effects \citep{Bassereh, Summhammer} but this should be investigated experimentally (or computationally, e.g. as an extension of the previous work \citep{Summhammer}) using sophisticated quantum molecular dynamics methods, since our results show that electrical interactions are much stronger than the filter backbone vibrations. 


\end{document}